\documentclass[journal]{IEEEtran}
\usepackage{amsmath,amsthm,amsfonts,amscd,amssymb,graphics,dsfont}
\usepackage[mathscr]{eucal}
\usepackage{graphics,graphicx,multicol}
\usepackage{epsfig}
\usepackage{epstopdf}
\usepackage{subfigure}
\usepackage{stfloats}
\usepackage{latexsym}
\usepackage{amsfonts}
\usepackage{cite}
\usepackage{algorithm,algorithmic}
\usepackage{color}

\begin{document}
\title{\huge{Deep Learning-based Limited Feedback Designs\\for MIMO Systems}}
\author{Jeonghyeon Jang, Hoon Lee,~\IEEEmembership{Member,~IEEE}, Sangwon Hwang, Haibao Ren, and Inkyu Lee,~\IEEEmembership{Fellow,~IEEE}
\thanks{J. Jang, S. Hwang, and I. Lee are with the School of Electrical Engineering, Korea University, Seoul, Korea (e-mail: $\{$march$\_$19, tkddnjs3510, inkyu$\}$@korea.ac.kr). 

H. Lee is with the Department of Information and Communications Engineering, Pukyong National University, Busan, Korea (e-mail: hlee@pknu.ac.kr).

H. Ren is with the Huawei Technologies, Shanghai, China (renhaibao@huawei.com).)}

}\maketitle 
\begin{abstract}
We study a deep learning (DL) based limited feedback methods for multi-antenna systems. Deep neural networks (DNNs) are introduced to replace an end-to-end limited feedback procedure including pilot-aided channel training process, channel codebook design, and beamforming vector selection. The DNNs are trained to yield binary feedback information as well as an efficient beamforming vector which maximizes the effective channel gain. Compared to conventional limited feedback schemes, the proposed DL method shows an 1 dB symbol error rate (SER) gain with reduced computational complexity.
\end{abstract}

\section{Introduction}
In the past decades, limited feedback techniques have been intensively investigated for realizing closed-loop communications in frequency-division duplex (FDD) systems \cite{Love:03,Clerckx:08,Xia:06, Lee:07, Kim:08, Kim:11}. A typical limited feedback procedure is divided into performing channel estimation and feeding back the information. Channel state information (CSI) is first estimated at the receiver \cite{Biguesh:06}, and then is mapped to the quantized values, e.g., the precoding matrix index (PMI) \cite{Love:03}. The quantized CSI is sent back to the transmitter through finite-rate feedback channels with the aid of a channel codebook \cite{Xia:06}.

A joint design of the channel estimation and quantization is necessary for identifying the optimal limited feedback systems, since the CSI estimation error is closely related to the codebook design. Nevertheless, due to non-convex and combinatorial nature of problem formulations, most works on the limited feedback systems \cite{Love:03,Clerckx:08,Lee:19} have studied the CSI estimation procedure and the quantization separately by assuming an ideal scenario where perfect CSI is available at the receiver. Another problem is a complexity issue on the channel estimation. It is expected that the estimation overhead grows as the training time increases.

Recently, a deep learning (DL) method has been applied for tackling limited feedback systems design problems \cite{Wen:18,Wang:18,Lu:18,Jang:19}. \cite{Wen:18} utilized sparsity of the massive multiple-input multiple-output (MIMO) channel using a convolutional neural network (CNN), and showed improved performance with lower complexity than other baseline schemes. A feedback method in a time-varying channel was addressed in \cite{Wang:18}, which adopted a recurrent neural network (RNN) to capture time-varying characteristic of the channel. The authors in \cite{Lu:18} deployed an integrated neural network composed of different neural network structures in both transmitter and receiver. In \cite{Jang:19}, feedback delay and error were considered in neural network designs. However, these works assumed perfect CSI and did not consider channel estimation. Therefore, a practical feedback system design is needed in the DL approach which takes the channel estimation process into account.

In this paper, we consider a point-to-point MIMO system where a multi-antenna receiver sends the quantized CSI back to a multi-antenna transmitter via a feedback channel. To this end, pilot sequences are first conveyed from the transmitter so that the receiver can extract useful features of the CSI. Then, we jointly design DL based limited feedback systems which include CSI prediction as well as codebook optimization. Two individual DNNs are implemented at the receiver and transmitter. The receiver DNN accepts the pilot-aided received signal as an input and is designed to output bipolar vectors as a quantized representation of the CSI. Such a DNN structure abstracts the channel estimation by directly extracting the binary feature from the received signal which contains the channel information. Also, the transmitter DNN is developed for calculating the beamforming vector for data transmission using the feedback information from the receiver \cite{Love:03}. We jointly train the DNNs at the transmitter and the receiver in an end-to-end manner so that the overall DL-based limited feedback scheme can learn efficient feedback rules by exploiting statistics of wireless channels. Consequently, compared to existing DL methods in \cite{Wen:18} where perfect CSI is assumed at the receiver, our proposed design is applicable to a more practical scenario with no exact CSI.

A major challenge for such a DNN training stems from the vector quantization operation at the receiver DNN whose gradient becomes zero for all input range. For this reason, gradient decent (GD) based DL libraries such as Tensorflow cannot be straightforwardly applied to the training task of the proposed DNN approach. It should be noted that the quantization process has not been included in the conventional DL studies \cite{Wen:18,Wang:18} as it is not easy to train DNNs with binary constraints. To address this issue, we employ the concept of a stochastic binarization layer and gradient estimation techniques \cite{Raiko:15}. As a result, the end-to-end training of the proposed DNN-based limited feedback system is made possible with state-of-the-art DL libraries. The numerical results verify that the proposed DL method improves the symbol error rate (SER) performance by 1 dB over conventional schemes with reduced computational complexity.

\section{System Model} \label{sec:main1}
\subsection{Limited Feedback Systems}

We consider a FDD MIMO system where a transmitter with $N_t$ antennas conveys the symbol $x$ to a receiver with $N_r$ antennas through the quasi-static frequency-flat fading channel $\mathbf{H}\in\mathbb{C}^{N_{r}\times N_{t}}$. In data transmission, denoting $\mathbf{w}\in\mathbb{C}^{N_{t}\times 1}$ as the beamforming vector, the received signal $\mathbf{y}^{\text{data}}\in\mathbb{C}^{N_{r}\times 1}$ is written by
\begin{align}
\mathbf{y}^{\text{data}} = \sqrt{E_{s}}\mathbf{H}\mathbf{s} + \mathbf{n}^{\text{data}} \label{eq:y}
\end{align}
where the precoded signal $\mathbf{s}\in\mathbb{C}^{N_t\times1}$ is given by $\mathbf{s}=\mathbf{w}x$, $E_{s}$ stands for the symbol transmission energy, and $\mathbf{n}^{\text{data}}\in\mathbb{C}^{N_{r}\times1}\sim\mathcal{CN}(\mathbf{0},\sigma^{2}_{n}\mathbf{I}_{N_{r}})$ is the additive Gaussian noise.

To achieve closed-loop communication, the receiver first estimates the channel matrix $\mathbf{H}$ using the standard pilot-based channel estimation methods \cite{Biguesh:06}. Defining $\mathbf{p}_{l}\in\mathbb{C}^{N_{t}\times1}$ ($\l=1,\cdots,L$) as the $l$-th pilot sequence, the received signal is given by $\mathbf{y}_{l}^{\text{train}}=\sqrt{E_p}\mathbf{H}\mathbf{p}_{l}+\mathbf{n}_{l}$, where $\mathbf{n}_{l}\in\mathbb{C}^{N_{r}\times 1}\sim\mathcal{CN}(\mathbf{0},\sigma_{n}^{2}\mathbf{I}_{N_{r}})$ accounts for the Gaussian noise vector. By stacking the received signals $\mathbf{y}_{l}^{\text{train}}$ for $l=1,\cdots,L$ into the matrix $\mathbf{Y}^{\text{train}}\triangleq[\mathbf{y}_{1}^{\text{train}},\cdots,\mathbf{y}_{L}^{\text{train}}]\in\mathbb{C}^{N_{r}\times L}$, we have
\begin{align}
\mathbf{Y}^{\text{train}}=\sqrt{E_p}\mathbf{HP}+\mathbf{N},\label{eq:Y}
\end{align}
where $E_p$ represents the energy for the pilot, and $\mathbf{P}$ and $\mathbf{N}$ are denoted as $\mathbf{P}\triangleq[\mathbf{p}_{1},\cdots,\mathbf{p}_{L}]\in\mathbb{C}^{N_{t}\times L}$ and $\mathbf{N}\triangleq[\mathbf{n}_{1},\cdots,\mathbf{n}_{L}]\in\mathbb{C}^{N_{r}\times L}$ respectively. The pilot matrix $\mathbf{P}$ is determined as the normalized discrete Fourier transform (DFT) matrix where $\mathbf{p}_i$ is defined as $\mathbf{p}_i \triangleq \frac{1}{\sqrt{N_t}}[1, e^{j2\pi(i-1)/L}, \cdots, e^{j2\pi(i-1)(N_t-1)/L} ]^{T}$. From \eqref{eq:Y}, the receiver can obtain the estimation $\hat{\mathbf{H}}$ of the CSI $\mathbf{H}$ by adopting the linear minimum mean square error (LMMSE) estimation~\cite{Biguesh:06}.

With the estimated CSI at hand, the receiver identifies the PMI by selecting a codeword in a pre-designed codebook $\mathcal{C}\triangleq\{\mathbf{w}_{1},\cdots,\mathbf{w}_{2^{B}}\}$ of size $2^{B}$, where $B$ stands for the number of feedback bits and $\mathbf{w}_{i}\in\mathbb{C}^{N_{t}\times 1}$ ($i=1,\cdots,2^{B}$) with $\|\mathbf{w}_{i}\|=1$ denotes the $i$-th candidate for the PMI. The receiver chooses the optimal PMI which maximizes the effective channel gain evaluated over $\hat{\mathbf{H}}$ as \cite{Love:03,Xia:06}
\begin{align}
i^{\star}=\arg\max_{i\in\{1,\cdots, 2^B\}}||\hat{\mathbf{H}}\mathbf{w}_{i}||_{2}^{2}. \label{eq:opt_PMI}
\end{align}
Here, the optimal codebook $\mathcal{C}$ maximizing the average effective channel gain $\mathbb{E}_{\mathbf{H}}[||\hat{\mathbf{H}}\mathbf{w}||_{2}^{2}]$ can be determined by the Lloyd algorithm \cite{Xia:06}. The receiver informs the index $i^{\star}$ to the transmitter through the feedback channel. Hence, the transmitter readily recovers the PMI based on the codebook, and the corresponding PMI is utilized as a beamforming vector for the data transmission over the channel $\mathbf{H}$.

In the limited feedback system, the receiver operation can be characterized as a mapping $i^{\star}=f_{R}(\mathbf{Y}^{\text{train}})$ which extracts the integer $i^{\star}$ from the received signal $\mathbf{Y}^{\text{train}}$. Similarly, the operation at the transmitter is generally represented by a function $\mathbf{w}=f_{T}(i^{\star})$ which calculates the beamforming vector $\mathbf{w}$ from the feedback information $i^{\star}$. Thus, an end-to-end limited feedback procedure can be written as $\mathbf{w} = f_{T}(f_{R}(\mathbf{Y}^{\text{train}}))$. An optimization task for the limited feedback scheme, which maximizes the effective channel gain over an arbitrarily distributed $\mathbf{H}$ and $\mathbf{N}$ in \eqref{eq:Y}, can be written by
\begin{align}
&\max\limits_{f_{T}(\cdot),f_{R}(\cdot)}\mathbb{E}_{\mathbf{H},\mathbf{N}}[||\mathbf{H}f_{T}(i^{\star})||_{2}^{2}]\label{eq:P1}\\[-2pt]
&\text{subject to } i^{\star}=f_{R}(\mathbf{Y}^{\text{train}})\in\{1,\cdots,2^{B}\}.\label{eq:C1}
\end{align}
Problem \eqref{eq:P1} handles an end-to-end optimization of the overall limited feedback process including the PMI extraction $i^{\star}=f_{R}(\mathbf{Y}^{\text{train}})$ and the beamforming vector computation $\mathbf{w} = f_{T}(f_{R}(\mathbf{Y}^{\text{train}}))$. The CSI estimation is abstracted in \eqref{eq:P1}, since the receiver does not explicitly predict the CSI $\mathbf{H}$ but obtains the quantization index $i^{\star}$ containing implicit information of~$\mathbf{H}$.

Existing codebook designs \cite{Love:03,Xia:06} developed for quantizing the estimated CSI cannot be straightforwardly applied to \eqref{eq:P1}, as a nontrivial feature is required to extract from the pilot-aided received signal $\mathbf{Y}^{\text{train}}$. In addition, it is difficult to solve \eqref{eq:P1} through traditional optimization methods, since closed-form expressions for the optimization functions $f_{T}(\cdot)$ and $f_{R}(\cdot)$ as well as the objective are not available for an arbitrary distributed channel $\mathbf{H}$. Hence, there is no general optimization approach for obtaining an efficient solution to \eqref{eq:P1}. To tackle this difficulty, we present a data-driven solution for the limited feedback systems through DL techniques.

\subsection{Basics of DNN}
Denoting $K_m$ as the dimension of the $m$-th hidden layer, the $m$-th hidden layer output $\mathbf{x}_m\in\mathbb{R}^{K_m \times 1}$ in a  fully-connected DNN with $M$ layers is expressed as
\begin{align}
\mathbf{x}_{m}=a_{m}(\mathbf{W}_{m}\mathbf{x}_{m-1}+\mathbf{o}_{m}),\label{eq:x_m}
\end{align}
where an element-wise function $a_{m}(\cdot)$ is defined as the activation function, and $\mathbf{W}_{m}\in\mathbb{R}^{K_{m}\times K_{m-1}}$ and $\mathbf{o}_{m}\in\mathbb{R}^{K_{m}\times 1}$ are a weight matrix and a bias vector of hidden layer $m$, respectively. Then, the overall DNN operation can be represented by a mapping $\mathbf{x}_{M+1}=g(\mathbf{x}_{0};\theta)$, which is composed of $M+1$ consecutive calculation in \eqref{eq:x_m}. The parameter set $\theta$ of the DNN is a collection of the weight matrices and the bias vectors, i.e., $\theta=\{\mathbf{W}_{m},\mathbf{b}_{m} \forall m\}$. The objective of the DNN training is to find the parameter $\theta$ minimizing a cost function, which mathematically describes the target of a DL task. State-of-the-art DL libraries such as Tensorflow depend on the gradient descent method and its variant for iteratively updating the DNN parameter $\theta$.

\section{Limited Feedback Systems Based on DL}
\begin{figure}
\begin{center}
\includegraphics[width=3.3in]{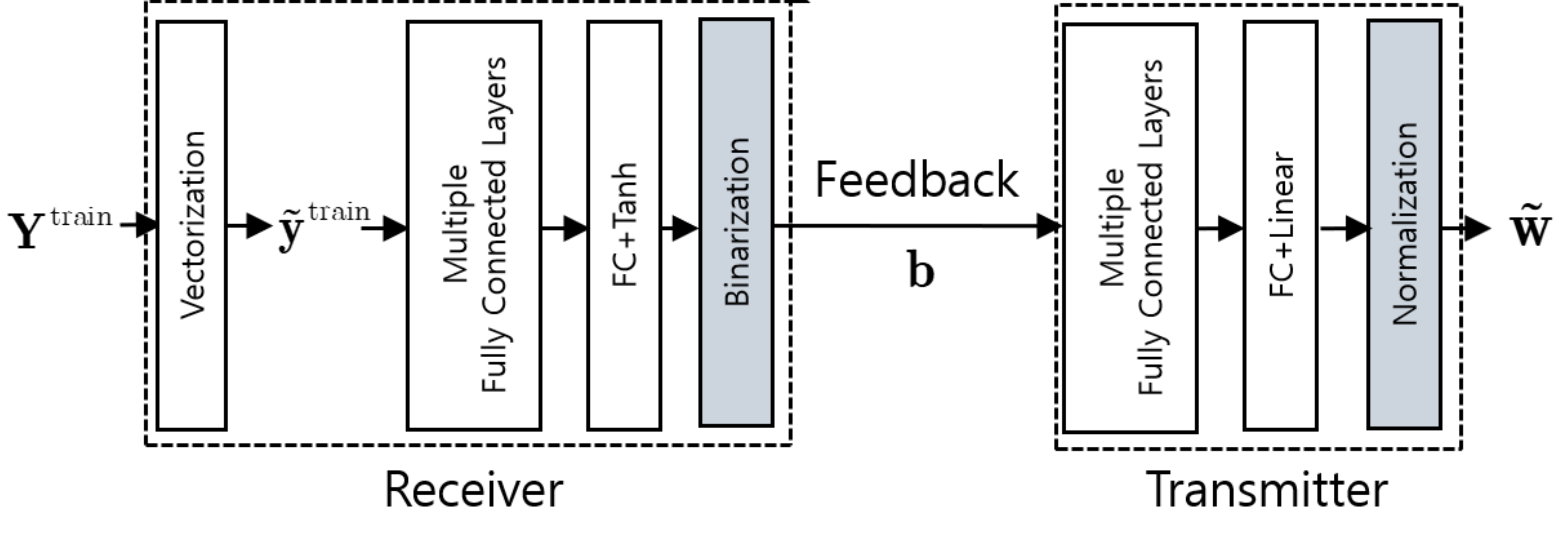}
\end{center}
\caption{Schematic diagram of DL-based limited feedback systems}
\label{figure:system_model}
\end{figure}

We propose a DL framework for the limited feedback systems illustrated in Fig. \ref{figure:system_model}. We employ two individual DNNs implemented at the receiver and the transmitter whose mapping is defined as $g_R(\cdot)$ and $g_T(\cdot)$, respectively, each of which approximates the unknown mappings $f_{R}(\mathbf{Y}^{\text{train}})$ and $f_{T}(i^{\star})$ of the overall limited feedback scheme.\footnote{The approximation accuracy of DNNs has been mathematically demonstrated both for continuous-valued functions \cite{KHornik:89} and discrete mappings \cite{ZLu:17}.} We design the receiver DNN such that it produces a bipolar vector $\mathbf{b}$ of length $B$ whose element is either $-1$ or $1$. The vector $\mathbf{b}$ is regarded as an equivalent representation of index $i^{\star}$ for the feedback information. The transmitter DNN constructs the unit-norm beamforming vector $\mathbf{w}$ maximizing the effective channel gain in \eqref{eq:P1}. In the following, we detail the operations of the proposed DL approach.

\subsection{Receiver}\label{sec:sec3.1}
At the receiver, the received signal matrix $\mathbf{Y}^{\text{train}}$ is first converted into the real vector representation $\tilde{\mathbf{y}}^{\text{train}}\triangleq[\Re\{\mathbf{y}^{\text{train}}\}\ \Im\{\mathbf{y}^{\text{train}}\}]\in\mathbb{R}^{1\times2LN_r}$ where $\mathbf{y}^{\text{train}}\triangleq\text{vec}(\mathbf{Y}^{\text{train}})$ stands for the vectorization of $\mathbf{Y}^{\text{train}}$, and $\Re\{\cdot\}$ and $\Im\{\cdot\}$ denote real and imaginary parts, respectively.\footnote{Tensorflow does not support complex number calculations.} It is then followed by a receiver DNN $\mathbf{b}=g_{R}(\tilde{\mathbf{y}}^{\text{train}};\theta_{R})$ parameterized by $\theta_{R}$ that accepts the received signal in \eqref{eq:Y} as an input and yields the bipolar vector $\mathbf{b}$ for the feedback information. We construct the receiver DNN with $M_{R}$ fully-connected hidden layers equipped with a rectified linear unit (ReLU) activation \cite{Lecun:15}, i.e., $a_{m}(z)\triangleq\max\{0,z\}$ for $m=1,\cdots,M_{R}$.

The dimension of the output layer is fixed to $B$ for generating $2^B$ different feedback values. To obtain the bipolar output, the activation at the output layer should be carefully chosen, since the quantization operation typically has zero gradient for all input range. This poses a {\em vanishing gradient} issue in the DNN training strategy where GD-based DL optimizers, e.g., the Adam algorithm \cite{Kingma:15}, fail to get updated and converge to a poor solution. To tackle this problem, we employ a stochastic binarization layer \cite{Raiko:15} which consists of two sequential activations. First, $\text{tanh}(z)\triangleq \frac{\text{exp}(2z)+1}{\text{exp}(2z)-1}$ is adopted to force the output of the receiver DNN within~a~range~$[-1,1]$.

Next, a stochastic activation $b(z)$, which adds a quantization noise $e$ to the output of $\text{tanh}(z)$, is chosen to mimic the quantization procedure as
\begin{align}
b(z) = z + e, \ \mbox{for}  \ z\in[-1,1],\label{eq:Bin}
\end{align}
where a distribution of the noise $e$ is determined as
\begin{align}
e = \left\{ \begin{array}{rcl} 1-z, & \mbox{with probability} & \frac{1+z}{2} \\ -1-z, & \mbox{with probability} & \frac{1-z}{2} \end{array}\right..\label{eq:prob.}
\end{align}
It is obvious that regardless of the value of $z$, the output of the stochastic binarization $b(z)$ becomes either $1$ or $-1$. The setup in \eqref{eq:prob.} leads to the zero-mean property for the quantization noise $\mathbb{E}_{e}[e]=0$, making $b(z)$ an unbiased estimator for the soft value $z$, i.e., $\mathbb{E}_{e}[b(z)|z]=z$.

Due to the probabilistic operation, the forward propagation \eqref{eq:Bin} of the stochastic binarization layer has no closed-form gradient expression which is not applicable to the GD-based DL libraries. This can be solved by the gradient estimation technique \cite{Raiko:15} where the gradient $\nabla_{\theta_{R}}b(z)$ with respect to $\theta_{R}$ is approximated as $\nabla_{\theta_{R}}b(z)\simeq\nabla_{\theta_{R}}\mathbb{E}_{e}[b(z)|z]$, i.e., the gradient over the quantization noise $e$. Thanks to the fact $\mathbb{E}_{e}[b(z)|z]\!=\!z$, the gradient is calculated as
\begin{align}
\nabla_{\theta_{R}}b(z) &\simeq \nabla_{\theta_{R}}\mathbb{E}_{e}[z+e|z] =\nabla_{\theta_{R}}z.\label{eq:grad.}
\end{align}
The approximation in \eqref{eq:grad.} is valid when the DNN experiences a large number of the quantization noise $e$, i.e., when the training set is sufficiently large. Note that the gradient estimation \eqref{eq:grad.} is only performed for the training where the gradient is computed via the backpropagation algorithm \cite{Lecun:15}. The forward propagation is produced with the stochastic binarization \eqref{eq:Bin}.

\subsection{Transmitter}
At the transmitter, we employ a DNN $\mathbf{w}=g_{T}(\mathbf{b};\theta_{T})$ which consists of $M_T$ hidden layers with the ReLU activations. The output layer of the transmitter DNN is implemented with the normalization activation $a_{M_{T}+1}(\mathbf{z})=\mathbf{z}/||\mathbf{z}||_{2}$ to produce a unit-norm beamforming vector. We denote the output of the transmitter DNN as $\tilde{\mathbf{w}}=[\Re\{\mathbf{w}\}\ \Im\{\mathbf{w}\}]\in\mathbb{R}^{2N_{t}\times1}$. Finally, the complex beamforming vector $\mathbf{w}$ is readily attained from converting the output $\tilde{\mathbf{w}}$.

\subsection{Training and Implementation}
We train the receiver DNN $g_{R}(\tilde{\mathbf{y}}^{\text{train}};\theta_{R})$ along with the transmitter DNN $g_{T}(\mathbf{b};\theta_{T})$ for the end-to-end optimization of the overall limited feedback procedure. To this end, we reformulate the original formulation in \eqref{eq:P1} as a training task of the DNNs by replacing the receiver and transmitter operations $f_{R}(\cdot)$ and $f_{T}(\cdot)$ with the DNNs $g_{R}(\tilde{\mathbf{y}}^{\text{train}};\theta_{R})$ and $g_{T}(\mathbf{b};\theta_{T})$, respectively. Thanks to the binarization layer given in Sec. \ref{sec:sec3.1}, the combinatorial constraint in \eqref{eq:C1} can be removed. We thus obtain the training problem as
\begin{align}
&\max\limits_{\theta_{T},\theta_{R}}\mathbb{E}_{\mathbf{H},\mathbf{N}}[||\mathbf{H}g_{T}(g_{R}(\tilde{\mathbf{y}}^{\text{train}};\theta_{R});\theta_{T})||_{2}^{2}],\label{eq:P2}
\end{align}
where the optimization variables now turn out to be the DNN parameters $\theta_{T}$ and $\theta_{R}$.

The training task \eqref{eq:P2} can be tackled via the mini-batch stochastic GD (SGD) algorithm \cite{Lecun:15}, which replaces the expectation in \eqref{eq:P2} with the empirical average over a mini-batch set $\mathcal{B}$ containing several samples of the training data. In our case, the training data set is composed of numerous tuples of CSI and the noise $(\mathbf{H},\mathbf{N})$ for generating the received signal matrix $\mathbf{Y}^{\text{train}}$ as the input to the DNNs. Defining $\Theta\!\!\triangleq\!\!\{\theta_{T},\theta_{R}\}$ as a collection of the DNN parameters, an iterative update rule at the $q$-th iteration of the mini-batch SGD is given~by
\begin{align}
\Theta^{[q]}\!\!=\!\!\Theta^{[q-1]}\!\!+\!\!\frac{\eta}{|\mathcal{B}|}\!\!\sum_{(\mathbf{H},\mathbf{N})\in\mathcal{B}}
    \!\!\!\!\!\!\nabla||\mathbf{H}g_{T}(g_{R}(\tilde{\mathbf{y}}^{\text{train}};\theta_{R});\theta_{T})||_{2}^{2},\label{eq:SGD}
\end{align}
where $\Theta^{[q]}$ represents the DNN parameter computed at the $q$-th iteration and $\eta>0$ is a learning rate.

The DNN training \eqref{eq:SGD} is an offline process, whereas the online computations of the trained DNNs are realized by simple linear matrix multiplications in \eqref{eq:x_m}. Once the DNNs are trained, we store the learned parameters $\theta_{T}$ and $\theta_{R}$ at the memory units of the transmitter and the receiver, respectively, for real-time limited feedback tasks. This can be seen as the sharing of the pre-designed PMI codebook in the conventional limited feedback scenarios. Notice that since the trained transmitter DNN $\mathbf{w}=g_{T}(\mathbf{b};\theta_{T})$ leads to an one-to-one mapping from the feedback information $\mathbf{b}$ to the corresponding beamforming vector $\mathbf{w}$, a lookup table implementation is possible for the transmitter. Thus, the computational complexity of the proposed DNN-based limited feedback scheme is dominated by the structure of the receiver DNN such as the dimension of each output of layer $(K_1^R,\cdots, K_{M_R+1}^R)$ and the number of hidden layers, and it is expressed as $\mathcal{O}(LN_{r}K_1^R+\sum_{m=1}^{M_{R}}K_m^RK_{m+1}^R+K_{M_R+1}^RB)$.

\section{Numerical Results} \label{sec:simulation}
In this section, we present numerical results for evaluating the performance of the proposed DL-based limited feedback schemes. Each element of $\mathbf{H}$ follows a zero-mean complex Gaussian distribution with covariance $\mathbf{R}_{\mathbf{H}}=\mathbb{E}_{\mathbf{H}}[\mathbf{H}^H\mathbf{H}]$.\footnote{Thanks to the data-driven training rule in \eqref{eq:SGD}, the proposed DL approach can be applied to any channel distribution.} Here, the $(i,j)$-th element $[\mathbf{R}_{\mathbf{H}}]_{i,j}$ of $\mathbf{R}_{\mathbf{H}}$ is fixed as \cite{Clerckx:08}
\begin{align}
[\mathbf{R}_{\mathbf{H}}]_{i,j} = \left\{ \begin{array}{rcl} N_rt^{|i-j|} & \mbox{for $i$}<\mbox{$j$} \\ N_r(t^*)^{|i-j|} & \mbox{otherwise}\end{array}\right.
\end{align}
where $t$ is the complex correlation coefficient. All the simulation results are averaged over the phase $\psi$ of $t$.

The receiver DNN consists of $M_{R}=3$ hidden layers each of which has the dimension $50N_{t}$, $30N_{t}$ and $20N_{t}$, while the transmitter DNN employs a reversed structure with respect to the receiver DNN with $M_{T}=3$ hidden layers whose output dimension is $20N_{t}$, $30N_{t}$ and $50N_{t}$. We randomly generate $|\mathcal{B}|=2000$ mini-batch samples for each SGD training iteration \eqref{eq:SGD} with $\eta=10^{-3}$, whereas for testing learned DNNs, we evaluate 100,000 independently sampled data. All the simulation is implemented with Tensorflow and Python.

\begin{figure}
\begin{center}
\includegraphics[width=3in]{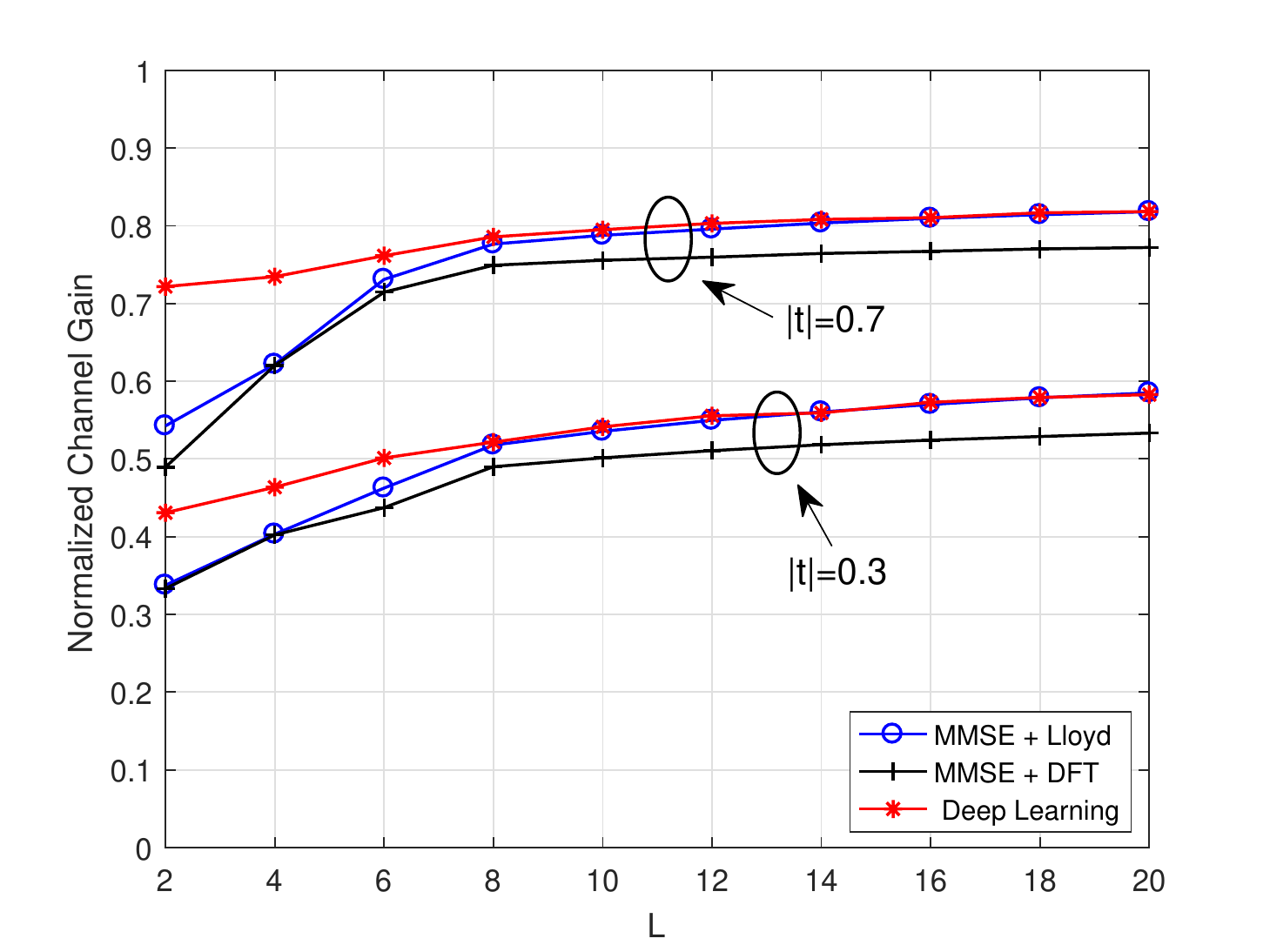}
\end{center}
\caption{Normalized channel gain with respect to $L$ with $N_{t}=8$, $N_{r}=4$~and~$B=6$.}
\label{figure:plot_L}
\end{figure}

Fig. \ref{figure:plot_L} illustrates the average normalized effective channel gain $\mathbb{E}_{\mathbf{H}}\left[||\mathbf{H}\mathbf{w}||^2/\lambda_{max}\right]$ of \eqref{eq:y} with $N_t=8$, $N_r=4$, $B=6$ as a function of the pilot sequence length $L$ for different correlation coefficients $t$ and the signal-to-noise ratio (SNR) defined as $\text{SNR}\triangleq\frac{E_{s}}{\sigma_{n}^{2}}$ where $\lambda_{max}$ denotes the maximum eigenvalue of $\mathbf{H}$. As a reference, we compare the performance of the proposed DL approach with conventional limited feedback systems which adopt the LMMSE channel estimation \cite{Biguesh:06} and the DFT channel codebook. We also examine the performance of the Lloyd algorithm which is known to be the optimal codebook design strategy under the assumption of no estimation error \cite{Xia:06}. For all schemes, we set $E_p/\sigma_n^2 = 0$ dB for the pilot transmission. From the figure, it is observed that the proposed DL method performs better than baseline schemes, especially when $L<N_t$. This implies that the proposed DL approach is more beneficial if the channel training duration is not long enough. It is due to sub-optimality which stems from separate optimization of the estimation and feedback process. We can also see that the DL-based limited feedback scheme offers a gain over the Lloyd codebook, which would not be optimal in the presence of the CSI estimation error.

\begin{figure}
\centering
\includegraphics[width=3in]{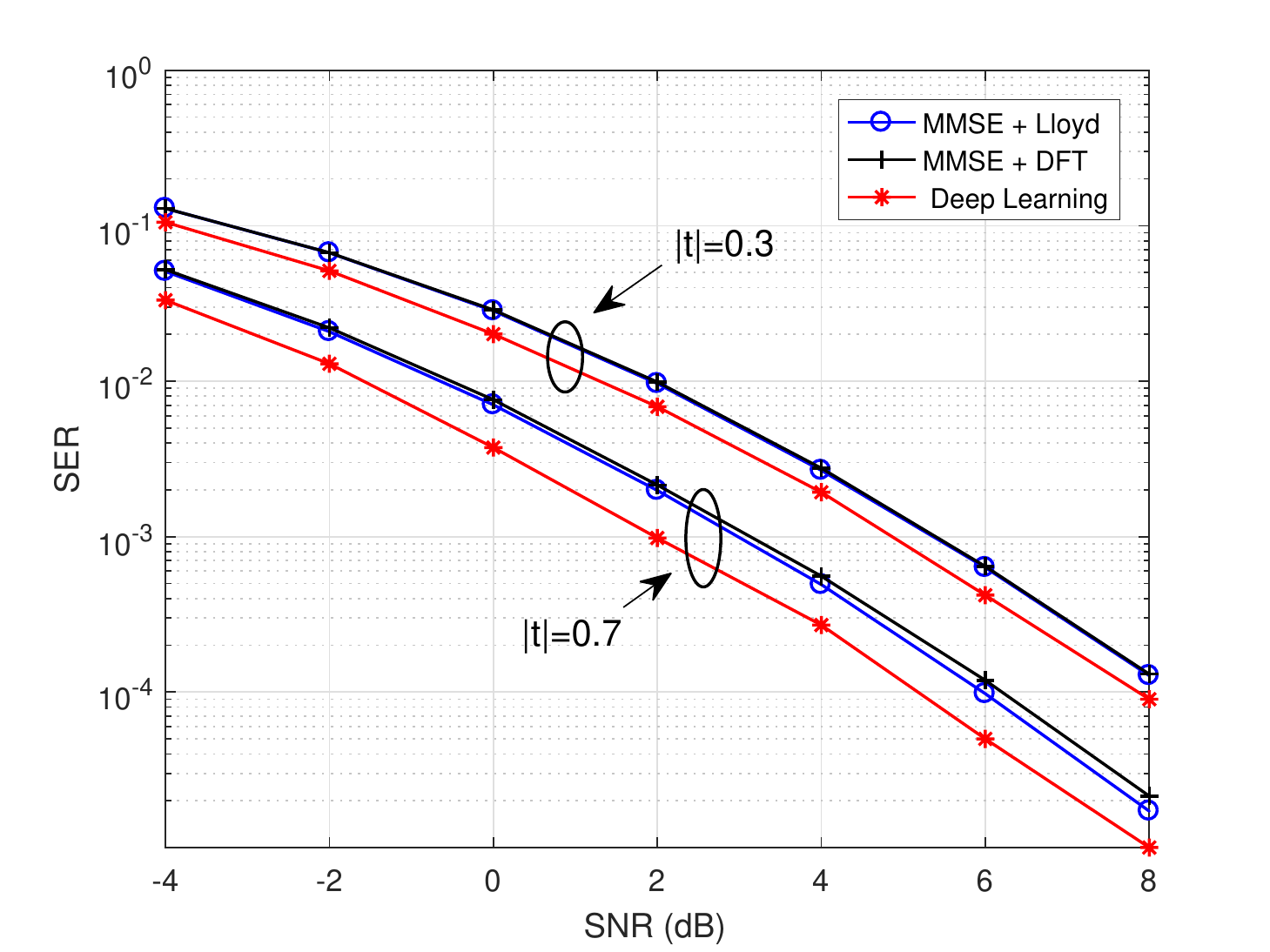}
\caption{SER with respect to SNR with $N_{t}=8$, $N_r=4$, $B=6$ and $L=4$.}
\label{figure:plot_SNR}
\end{figure}

We investigate the average symbol error rate (SER) performance in Fig. \ref{figure:plot_SNR} for the QPSK modulated systems with $L=4$. It is clear that our proposed method exhibits lower SER compared to baseline schemes. Also, the performance of DL is shown to be more effective when the antennas are highly correlated. It can be shown that proposed schemes offers about an 1 dB gain over the other methods for $|t|=0.7$.

\begin{table}[]
\centering
\caption{Comparison of the CPU running time [$\mathrm{msec}$]}
\begin{tabular}{|c||c|c|c|c|}
\hline
                  & L=2    & L=5    & L=10   & L=20   \\ \hline\hline
MMSE + Lloyd, DFT & 0.0625 & 0.0668 & 0.0771 & 0.0981 \\ \hline
Deep Learning     & 0.0270 & 0.0286 & 0.0290 & 0.0311 \\ \hline
\end{tabular}
\label{table:CPU Time}
\end{table}

Finally, we compare the average CPU running time in Table \ref{table:CPU Time}. For fair comparison, the execution time of the baseline methods only includes the online computations, i.e., the channel estimation via the LMMSE and the quantization \eqref{eq:opt_PMI}. Since the Lloyd algorithm is an offline procedure, two baselines have the identical complexity. For the DL approach, the real-time complexity is only rely on the receiver DNN. We can see that the DL-based scheme reduces the execution time of the baseline methods by half. This verifies the effectiveness of the proposed DL approach for practical limited feedback designs.

\section{Conclusion} \label{sec:conclusion}
We have proposed DL-based limited feedback methods for the MIMO systems. The overall limited feedback process has been implemented by two DNNs at the receiver and the transmitter. Both DNNs have been jointly trained to produce efficient quantization and beamforming vectors. Numerical results have demonstrated that the proposed DL approach can improve the performance of conventional limited feedback schemes with reduced complexity.

\nocite{*}
\bibliography{arXiv}
\bibliographystyle{ieeetr}

\end{document}